\documentclass[aps,pre,amsmath,amssymb,reprint,groupedaddress,showpacs]{revtex4-1}

\usepackage{graphicx}
\usepackage{dcolumn}
\usepackage{bm}

\begin{document}


\title{Electromagnetic space-time crystals. II. Fractal computational approach}


\author{G. N. Borzdov}
\email[]{BorzdovG@bsu.by}
\affiliation{Department of Theoretical Physics and Astrophysics, Belarus State University,
Nezavisimosti avenue 4, 220030 Minsk, Belarus}



\begin{abstract}
A fractal approach to numerical analysis of electromagnetic space-time crystals, created by three standing plane harmonic waves with mutually orthogonal phase planes and the same frequency, is presented. Finite models of electromagnetic crystals are introduced, which make possible to obtain various approximate solutions of the Dirac equation. A criterion for evaluating accuracy of these approximate solutions is suggested.
\end{abstract}


\maketitle

\section{Introduction}
Fractal approach makes possible to obtain and/or research objects with any level of complexity by using simple algorithms~\cite{Falc}. It provides useful tools to design noval devices, such as fractal antennas, filters, diffusers, absorbes, microwave invisibility cloacks, and fractal metamaterials~\cite{Takeda,BoHou,Krzy}, as well as effective algorithms for computer qraphics and fractal compression of digital images~\cite{image}.

In this paper we present a fractal computational approach to calculating the fundamental solution of the Dirac equation describing the motion of an electron in an electromagnetic field with four-dimensional (4d) periodicity (electromagnetic space-time crystal, or ESTC)~\cite{ESTCp1}. The electromagnetic field is composed of three standing plane harmonic waves with mutually orthogonal phase planes and the same frequency. In  this case, the Dirac equation reduces to an infinite system of linear matrix equations. Each equation of the system relates 13 Fourier amplitudes [bispinors $c(n+s)$], where the multi-index $n=(n_1,n_2,n_3,n_4)$ is a point of the integer lattice $\mathcal{L}$ with even values of the sum $n_1+n_2+n_3+n_4$, and the shift $s=(s_1,s_2,s_3,s_4)\in \mathcal{L}$ takes all 13 values satisfying the condition $g_{4d}(s)=0, 1$, by definition (see appendix) $g_{4d}(s_1,s_2,s_3,s_4)=\max\{|s_1|+|s_2|+|s_3|,|s_4|\}$.

In other words, each amplitude $c(n)$ enters in 13 different matrix equations of the infinite system. The fundamental solution of this system is obtained in the previous paper~\cite{ESTCp1} by a recurrent process. It is expressed in terms of an infinite series of projection operators. This process begins with the selection of an infinite subsystem consisting from independent equations and the calculation of the projection operators $\rho_0(n)=P(n), \quad n\in \mathcal{F}_0 \subset \mathcal{L}$, which uniquely define the fundamental solutions of these equations~\cite{ESTCp1}. At each new step of the recurrent process, we add another infinite set of mutually independent equations (MIE) which, however, are related with some of the equations introduces at the previous steps. Consequently, we obtain an infinite set of independent finite systems of interrelated equations [fractal clusters of equations (FCE)]. It can be described as a 4d lattice of such clusters. Each step of the recurrent procedure expands FCE for which it provides the exact fundamental solutions. The presented in Sec.~\ref{sec:fractal} fractal algorithm of this expansion is devised to minimize volumes of computations and data files. Some MIE (aggregative MIE, or MIE1) just add one equation to each cluster of the previous FCE lattice so that these enlarged clusters remain independent. Other MIE (connective MIE, or MIE2), by adding each equation, interrelate a pair of neighboring clusters into a joint cluster, and a quite different FCE lattice arises. Each fractal period includes connections in directions of $n_4, n_1, n_2$, and $n_3$ axes, respectively. The smaller is FCE, the smaller are volumes of computations and data files, which are necessary to find and to write down the fundamental solution for this FCE. To simplify calculations, we add a maximal possible number of MIE1 before adding the next MIE2.

In Sec.~\ref{sec:approx}, we discus the interrelations between the fundamental solution and approximate partial solutions which can be obtained in the framework of finite models of ESTCs. In Sec.~\ref{sec:accur}, a criterion for evaluating accuracy of approximate solutions is suggested, which plays a great role in numerical analysis of ESTCs. Results of this analysis will be presented in the subsequent paper. The introduced in appendix sequential numbering of points $n\in \mathcal{L}$ drastically simplifies numerical implementation of the presented technique and analysis of solutions.

\section{\label{sec:fractal}Fractal splitting of the fundamental solution}
Due to the specific Fourier spectrum~\cite{ESTCp1} of the 4d periodic electromagnetic field of ESTC, in the current series of papers we use indexing of Fourier components of the wave function and many other mathematical objects by points $n=(n_1,n_2,n_3,n_4)$ of the integer lattice $\mathcal{L}$ with even values of the sum $n_1+n_2+n_3+n_4$.  The fundamental solution $\mathcal{S}$ and the projection operator $\mathcal{P}$ of the infinite system of equations under study are defined as follows~\cite{ESTCp1}
\begin{equation}\label{Pro}
  \mathcal{S}=\mathcal{U}-\mathcal{P},\quad  \mathcal{P}=\sum_{k=0}^{+\infty}\sum_{n\in\mathcal{F}_k}\rho_k(n),
\end{equation}
\begin{equation}\label{FkL}
    \bigcup_{k=0}^{+\infty}\mathcal{F}_k=\mathcal{L}, \quad \mathcal{F}_j\bigcap\mathcal{F}_k=\emptyset \quad j\neq k,
\end{equation}
where $\mathcal{U}$ is the unit operator and $\rho_k(n)$ are Hermitian projection operators with the trace $tr[\rho_k(n)]=4$. To specify the lattices $\mathcal{F}_k$, we split the lattice $\mathcal{L}$ into fractal subsets and designate stages of calculation by $(s_t,p_h)$, where $s_t=0,1,\dots$, and $p_h=1,2$ is the phase of stage $s_t$.

Let $\mathcal{L}_{set}[(a_1,a_2,a_3,a_4),(b_1,b_2,b_3,b_4)]$ be the subset of $\mathcal{L}$ given by $a_i\leq n_i\leq b_i, i=1,2,3,4$, where $a_i$ and $b_i$ are some integers. At the initial stage $s_t=0$, we split $\mathcal{L}$ into subsets $\mathcal{L}_{ss}[(0,1),s]\equiv \mathcal{L}_{ss}[(0,2),s]$ which can be obtained by periodic translation of the central subset
\begin{equation*}
    \mathcal{L}_{cs}(0,1)\equiv\mathcal{L}_{cs}(0,2)=\mathcal{L}_{set}[(-1,-1,-1,-1),( 2,2,2,2)]
\end{equation*}
using the list of periods $p_{st}(0)=\{4,4,4,4\}$, i.e., by shifts $s=4(k_1,k_2,k_3,k_4)$, where $k_i$ are integers. Here, $\mathcal{L}_{ss}[(s_t,p_h),s]$ signifies the subset obtained by a shift $s$ of the central subset $\mathcal{L}_{cs}(s_t,p_h)$ of stage $(s_t,p_h)$. Each subset $\mathcal{L}_{ss}[(0,1),s]$ contains 128 points of $\mathcal{L}$.

The lattice $\mathcal{L}$ can be composed of point lattices numbered $u=1,2,\dots$, and specified by the center $c_0(u)$ and the list of periods $p_0(u)$. At stage $(0,1)$, we compose the lattice $\mathcal{F}_0$ of 8 sublattices ($u=1,\dots,8$) with equal periods $p_0(u)=p_{st}(0)=\{4,4,4,4\}$ and the following list of centers
\begin{eqnarray}
  c_L(0,1)\equiv&&\{c_0(u),u=1,\dots,8\}\\
   =&&\left\{(0,0,0,0),(-1,-1,-1,-1),\right.\nonumber\\
   &&(1,1,-1,-1),(1,-1,1,-1),(-1,1,1,-1),\nonumber\\
   &&\left.(0,2,2,0),(2,0,2,0),(2,2,0,0)\right\}\subset\mathcal{L}_{cs}(0,1).\nonumber
\end{eqnarray}
It is easy to verify that, any two points $n, n' \in \mathcal{F}_0$ satisfy the condition $g_{4d}(s)>2$, where $s=(s_1,s_2,s_3,s_4)=n'- n$. In this case, $\rho_0(n)\rho_0(n')=0$~\cite{ESTCp1}.

At stage $(0,2)$, we introduce the next 6 lattices $\mathcal{F}_1,\dots,\mathcal{F}_6$ which have the same periods $p_0(u)=p_{st}(0)=\{4,4,4,4\}$, numbers $u=8+k=9,\dots,14$, and the list of centers
\begin{eqnarray}
  c_L(0,2)\equiv&&\{c_0(u),u=9,\dots,14\}\\
   =&&\left\{(0,0,-1,-1),(0,-1,0,-1),(-1,0,0,-1),\right.\nonumber\\
   &&\left.(0,1,1,0),(1,0,1,0),(1,1,0,0)\right\}\subset\mathcal{L}_{cs}(0,2).\nonumber
\end{eqnarray}
For any points $n$ and $n'$ of lattices with numbers $u$ and $u'$ ($1\leq u,u'\leq 14$), $n'-n=c_0(n')-c_0(n)+4\{k_1,k_2,k_3,k_4\}$, where $k_i$ are some integers. It is easy to check that $g_{4d}(n'-n)>2$ if, at least, one of $k_i$ is not zero. Because of this, one can calculate $\rho_k(m)$ at $k=1,\dots,6$ by taking into account only those $\rho_j(n)$ for which $j=0,\dots,k-1$ and $n$ belongs to the subset $\mathcal{L}_{ss}[(0,2),s]$ containing $m$. This conclusion follows immediately from the recurrent relations presented in~\cite{ESTCp1}.

We define lattices $\mathcal{F}_k $ in such a way that, at any stage $(s_t,p_h)$, calculations of $\rho_k(m)$ and $\rho_k(m')$ can be carried out independently at different subsets: $\mathcal{L}_{ss}[(s_t,p_h),s]$ and $\mathcal{L}_{ss}[(s_t,p_h),s']$. To fulfil this condition, one can add only a finite number of point lattices at each stage, in particular 8 and 6 at stages $(0,1)$ and $(0,2)$, respectively. Besides, at $k>0$, $\mathcal{F}_k$ comprises only point lattice with $u=8+k$. At stages $s_t=1,2,3,4$, which constitute the first cycle of fractal expansion, we have the following lists of periods:
\begin{eqnarray*}
 p_0(u)&=&p_{st}(1)=\{4,4,4,12\} \text{ for } u = 15,\dots,42,\\
       &=&p_{st}(2)=\{12,4,4,12\} \text{ for } u=43,\dots,102,\\
       &=&p_{st}(3)=\{12,12,4,12\} \text{ for } u=103,\dots,402,\\
       &=&p_{st}(4)=\{12,12,12,12\} \text{ for } u=403,\dots,2222,
\end{eqnarray*}
and the central subsets $\mathcal{L}_{cs}(s_t,p_h)$:
\begin{eqnarray*}
  \mathcal{L}_{cs}(1,1)&=&\mathcal{L}_{set}[(-1,-1,-1,-5),(2,2,2,2)],\\
  \mathcal{L}_{cs}(1,2)&=&\mathcal{L}_{set}[(-1,-1,-1,-5),(2,2,2,6)],
\end{eqnarray*}
\begin{eqnarray*}
  \mathcal{L}_{cs}(2,1)&=&\mathcal{L}_{set}[(-5,-1,-1,-5),(2,2,2,6)],\\
  \mathcal{L}_{cs}(2,2)&=&\mathcal{L}_{set}[(-5,-1,-1,-5),(6,2,2,6)],
\end{eqnarray*}
\begin{eqnarray*}
  \mathcal{L}_{cs}(3,1)&=&\mathcal{L}_{set}[(-5,-5,-1,-5),(6,2,2,6)],\\
  \mathcal{L}_{cs}(3,2)&=&\mathcal{L}_{set}[(-5,-5,-1,-5),(6,6,2,6)],
\end{eqnarray*}
\begin{eqnarray*}
  \mathcal{L}_{cs}(4,1)&=&\mathcal{L}_{set}[(-5,-5,-5,-5),(6,6,2,6)],\\
  \mathcal{L}_{cs}(4,2)&=&\mathcal{L}_{set}[(-5,-5,-5,-5),(6,6,6,6)].
\end{eqnarray*}

At the phases $p_h=1$ and $p_h=2$ of any stage $s_t$, equal numbers of point lattices are added, namely, 14, 30, 150, 910 for $s_t=1,2,3,4$, respectively. At $s_t=1$, the centers of lattices ($u=8+k=15,\dots,42$) are defined as
\begin{equation}\label{cun15}
    c_0(u+14p_h)=c_0(u)+(-1)^{p_h}(0,0,0,2),
\end{equation}
where $u= 1,\dots,14, p_h=1,2$. At $s_t=2,3$, and $4$, the centers of the lattices added at $p_h=1$ and $p_h=2$ are related by shifts as follows:
\begin{eqnarray}
  c_0(u+30)=c_0(u)+(4,0,0,0),\label{cun43}\\
   s_t=2, u=8+k=43,\dots,72;\nonumber
\end{eqnarray}
\begin{eqnarray}
  c_0(u+150)=c_0(u)+(0,4,0,0), \label{cun103}\\
   s_t=3, u=8+k=103,\dots,252;\nonumber
\end{eqnarray}
\begin{eqnarray}
  c_0(u+910)=c_0(u)+(0,0,4,0),\label{cun403}\\
   s_t=4, u=8+k=403,\dots,1312.\nonumber
\end{eqnarray}

To define $c_0(u)$ at $p_h=1$, we use the lists:
\begin{eqnarray*}
  c'_L(4,1)=\left\{(3, 5, 0), (5, 3, 0), (3, 4, -1), (5, 4, -1),\right.\\
   (6, 5, -1), (4,  3, -1), (4, 5, -1), (5, 5, -2),\\
   (5, 6, -1), (4, 5, -2), (5,  4, -2), (4, 4, -3),\\
   (3, 3, -2), (3, 4, -2), (4, 3, -2), (4,  4, -2),\\
    (3, 3, -3), (-1, 5, 0), (1, 3, 0), (-1, 4, -1),\\
   (1,  4, -1), (2, 5, -1), (0, 3, -1), (0, 5, -1),\\
   (1, 5, -2), (1,  6, -1), (0, 5, -2), (1, 4, -2),\\
   (2, 3, -2), (0, 4, -3), (2,  3, -1), (-1, 3, -2),\\
   (-1, 4, -2), (2, 4, -1), (0, 3, -2), (2, 4,  0),\\
   (0, 4, -2), (-1, 3, -3), (1, 3, -1), (-3, 3, 0),\\
   (-3,  4, -1), (-2, 5, -1), (-4, 3, -1), (-3, 5, -2),\\
   (-3, 6, -1), (-4,  5, -2), (-3, 4, -2), (-2, 3, -2),\\
   (-4, 4, -3), (-2, 3, -1), (-2,  4, -1), (-4, 3, -2),\\
   (-2, 4, 0), (-3, 3, -1), (3, 1, 0), (5, -1, 0),\\
   (3, 0, -1), (5, 0, -1), (6, 1, -1), (4, -1, -1),\\
   (4, 1, -1), (5,  1, -2), (5, 2, -1), (3, 2, -2),\\
   (4, 1, -2), (5, 0, -2), (4,  0, -3), (3, 2, -1),\\
   (3, -1, -2), (3, 0, -2), (4, -1, -2), (4,  2, -1),\\
   (4, 2, 0), (4, 0, -2), (3, -1, -3), (3, 1, -1),\\
   (-1, 1,  0), (1, -1, 0), (-1, 0, -1), (1, 0, -1),\\
   (2,  1, -1), (0, -1, -1), (0, 1, -1), (1, 1, -2),\\
   (1, 2, -1), (-1,  2, -2), (0, 1, -2), (1, 0, -2),\\
   (2, -1, -2), (0,  0, -3), (2, -1, -1), (-1, 2, -1),\\
   (2, 2, -3), (-1, -1, -2), (-1,  0, -2), (2, 1, -2),\\
   (2, 0, -1), (0, -1, -2), (1, 2, -2), (0,  2, -1),\\
   (2, 2, -2), (2, 0, 0), (0, 2, 0), (0,  0, -2),\\
   (-1, -1, -3), (1, 1, -3), (-1, 1, -1), (1, -1, -1),\\
   (-3, -1,  0), (-3, 0, -1), (-2, 1, -1), (-4, -1, -1),\\
   (-3, 1, -2), (-3,  2, -1), (-5, 2, -2), (-4, 1, -2),\\
   (-3, 0, -2),  (-2, -1, -2), (-4,  0, -3), (-2, -1, -1),\\
   (-2, 2, -3), (-2, 1, -2), (-2,  0, -1), (-4, -1, -2),\\
   (-3, 2, -2), (-4, 2, -1), (-2, 2, -2), (-2, 0,  0),\\
   (-3, 1, -3), (-3, -1, -1), (3, -3,  0), (3, -4, -1),\\
   (6, -3, -1), (4, -3, -1), (5, -3, -2), (5, -2, -1),\\
   (3, -2, -2), (4, -3, -2), (5, -4, -2), (4, -4, -3),\\
   (3, -2, -1), (3, -4, -2), (4, -2, -1), (4, -2, 0),\\
   (3, -3, -1), (-1, -3,  0), (-1, -4, -1), (2, -3, -1),\\
   (0, -3, -1), (1, -3, -2), (1, -2, -1), (-1, -2, -2),\\
   (0, -3, -2), (1, -4, -2), (2, -5, -2), (0, -4, -3),\\
   (-1, -2, -1), (2, -2, -3), (-1, -4, -2), (2, -3, -2),\\
   (2, -4, -1), (1, -2, -2), (0, -2, -1), (2, -2, -2),\\
   (0, -2,  0), (1, -3, -3), (-1, -3, -1), (-2, -3, -1),\\
   (-3, -3, -2), (-3, -2, -1), (-5, -2, -2), (-4, -3, -2),\\
   (-3, -4, -2), (-2, -5, -2), (-4, -4, -3), (-2, -2, -3),\\
   (-2, -3, -2), (-2, -4, -1), (-3, -2, -2), (-4, -2, -1),\\
   \left. (-2, -2, -2), (-3, -3, -3)\right\},
  \end{eqnarray*}
\begin{eqnarray*}
  c'_L(3,1)=\left\{(6, -1, 1), (3, -1, 0), (3, 0, 1), (5, -2, 1),\right.\\
   (5, -1, 2), (4, -1, 1),(4, -2, 1), (4, -3, 0),\\
   (5, -2, 0), (3, -2, 0), (2, -1,  1), (-1, -1, 0),\\
   (-1, 0, 1), (1, -2, 1), (1, -1, 2), (0, -1, 1),\\
   (0, -2, 1), (0, -3, 0), (1, -2, 0), (2, -2, -1),\\
   (-1, -2,  0), (2, -1, 0), (-2, -1, 1), (-3, -2, 1),\\
   (-3, -1, 2), (-4, -2, 1), (-4, -3, 0), (-3, -2, 0),\\
   \left.(-2, -2, -1), (-2, -1, 0)\right\},
\end{eqnarray*}
\begin{eqnarray*}
  c'_L(2,1)=\left\{(-1,2,1),(-1,1,2),(-2,1,1),\right.\\
  \left. (-2,0,1),(-3,0,0),(-2,1,0)\right\}.
\end{eqnarray*}
Each five consecutive centers $n=(n_1,n_2,n_3,n_4)=c_0(u)$ in any of Eqs.~(\ref{cun43})--(\ref{cun403}) have the same projection onto the three-dimensional ($3d$) space consisting of points $(n_1,n_2,n_3)$. They differ only by values of $n_4$, namely, $n_4=4, 0, -4, 2, -2$ if $|n_1|+|n_2|+|n_3|$ is even, and $n_4=3, -1, -5, 1, -3$ if $|n_1|+|n_2|+|n_3|$ is odd. In particular, $c_0(43)=(-1,2,1,4), c_0(44)=(-1,2,1,0),\dots, c_0(47)=(-1,2,1,-2),c_0(48)=(-1,1,2,4)$ and so on. Because of this, the presented above lists of 3d projections $c'_L(s_t,1)$ and Eqs.~(\ref{cun43})--(\ref{cun403}) uniquely define the centers $c_0(u)$ at $s_t=2, 3, 4$, and $u=43,\dots,2222$.

In a similar way the next fractal cycle runs through the eight stages $(5,1), (5,2),\dots,(8,2)$. It begins at $u=2223$  with
\begin{eqnarray*}
 c_0(2223)=(4,4,4,-6), p_{st}(5)=\{12,12,12,36\}, \\
  \mathcal{L}_{cs}(5,1)=\mathcal{L}_{set}[(-5,-5,-5,-17),(6,6,6,6)]
\end{eqnarray*}
and ends at $u=108526$ with
\begin{eqnarray*}
 c_0(108526)=(-15,-15,5,-7), p_{st}(8)=\{36,36,36,36\}, \\
  \mathcal{L}_{cs}(8,2)=\mathcal{L}_{set}[(-17,-17,-17,-17),(18,18,18,18)].
\end{eqnarray*}

\section{\label{sec:approx}Approximate solutions}
Numerical implementation of the obtained solution implies the replacement of the projection operator $\mathcal{P}$ (\ref{Pro}) of the infinite system of equations \cite{ESTCp1}
\begin{equation}\label{PnC}
    P(n)C = 0, \quad n \in \mathcal{L}
\end{equation}
by the projection operator
\begin{equation}\label{Pprime}
    \mathcal{P'}=\sum_{k\in k_L}\sum_{n\in n_L(k)}\rho_k(n)
\end{equation}
of its finite subsystem
\begin{equation}\label{PkbLC}
    P(n)C=0,\quad n\in \mathcal{L}'=\bigcup_{k\in k_L}n_L(k) \subset \mathcal{L},
\end{equation}
where $k_L$ is an ordered finite list of integers, and $n_L(k)$ is a finite list of points $n\in\mathcal{F}_k$, taking into account. Here, $C$ is the so-called multispinor~\cite{ESTCp1} defined as the set $C =\{c(n),n \in {\mathcal L}\}$ of the bispinor Fourier amplitudes $c(n)$ of the wave function in the Dirac equation, treated as an element of an infinite dimensional linear space $V_C$.

The projection operator
\begin{equation}\label{SkbL}
    \mathcal{S'} = \mathcal{U} - \mathcal{P'}
\end{equation}
defines the exact fundamental solution of Eq.~(\ref{PkbLC}) and an approximate solution of Eq.~(\ref{PnC}). In particular, using the described above fractal lattices $\mathcal{F}_k$, we can set
\begin{equation}\label{kL2214}
    k_L=\{0,1,\dots,2214\},\quad n_L(k)=\mathcal{F}_k\bigcap \mathcal{L}_{cs}(4,2).
\end{equation}
In this case, the system~(\ref{PkbLC}) contains 5150 equations, including all 2048 equations with $n\in \mathcal{L}_{set}[(-3,-3,-3,-3),(4,4,4,4)]\subset\mathcal{L}$. In some special applications, it may be advantageous to restrict both the recurrent relations \cite{ESTCp1} and the system (\ref{PkbLC}) to a subset of $k_L$~(\ref{kL2214}) and subsets of $n_L(k)$~(\ref{kL2214}), i.e., to a more simple finite model of the infinite electromagnetic crystal.

In the subsequent paper, we will illustrate the presented technique by some results of its computer simulation. To this end, we restrict our consideration to the case when the amplitude $C_0$ specifying a partial solution~\cite{ESTCp1} is given by
\begin{equation}\label{C0n0}
    C_0=a_0^j e_j(n_o),\quad n_o=(0,0,0,0),
\end{equation}
and $\mathcal{L}'\subset \mathcal{L}_{cs}(4,2)$. Here, $e_j(n)$ is the basis in $V_C$~\cite{ESTCp1}, and summation over repeated indices is carried out from 1 to 4. In this case, the relation
\begin{equation}\label{CSd}
    C=\{c(n),n\in S_d\}=\mathcal{S'}C_0=C_0-\mathcal{P'}C_0
\end{equation}
describes the four-dimensional subspace of exact solutions of Eq.~(\ref{PkbLC}), i.e., for any given bispinor
\begin{equation}\label{a0}
    a_0=\left(
           \begin{array}{c}
               a_0^1 \\
               a_0^2 \\
               a_0^3 \\
               a_0^4 \\
             \end{array}
             \right),
\end{equation}
it specifies a partial solution, where $S_d\subset\mathcal{L}$ is the subset of $\mathcal{L}$ with nonzero bispinors $c(n)$, for brevity sake, it will be referred as the solution domain.

Bisbinors $c(n)$ and $a_0$ are linearly related as
\begin{equation}\label{Snc0}
    c(n)=S(n)a_0,
\end{equation}
where $S(n)$ is the $4\times 4$ matrix of components $\left\langle\theta^i(n),\mathcal{S'}e_j(n_o)\right\rangle$ of the operator
\begin{equation}\label{InSIn0}
    I(n)\mathcal{S'}I(n_o)=S^i{}_j(n)e_i(n)\otimes\theta^j(n_o).
\end{equation}
Here, $\theta^j(n) = e_j^{\dag}(n)$ is the dual basis in the space of one-forms $V_C^\ast$,  $I(n)=e_j(n)\otimes\theta^j(n)$ is the projection operator related with point $n\in \mathcal{L}$. From Eqs.~(\ref{Pprime}), (\ref{SkbL}), and (\ref{InSIn0}) it follows
\begin{equation}\label{Sn}
    S(n)=U \delta(n-n_o) - \sum_{k\in k_L}\sum_{m\in n_L[k]}R_k(n,m,n_o),
\end{equation}
where $U$ is the unit $4\times 4$ matrix, $\delta(n-n_o)$ is the Kronecker delta, matrices $R_k(n,m,n_o)$ are defined in \cite{ESTCp1}. Substituting of $c(n)$ into the Fourier series, specifying the bispinor wave function $\Psi$~\cite{ESTCp1}, gives
\begin{equation}\label{Exc0}
    \Psi(\bm{x})=\sum_{n\in S_d}c(n)e^{i \varphi_n(\bm{x})}\equiv E(\bm{x})a_0,
\end{equation}
where $\bm{x}=(\textbf{r},ict)$, and
\begin{equation}\label{Ex}
    E(\bm{x})=\sum_{n\in S_d}e^{i \varphi_n(\bm{x})}S(n)
\end{equation}
is the evolution operator. In terms of the dimensionless coordinates $\textbf{r}'=\textbf{r}/\lambda_0=X_1\textbf{e}_1+X_2\textbf{e}_2+X_3\textbf{e}_3$, $X_4=ct/\lambda_0$, and the dimensionless parameters
\begin{equation}
    \textbf{q} = \frac{\hbar\textbf{k}}{m_e c},\quad q_4 = \frac{\hbar \omega}{m_e c^2},\quad \Omega = \frac{\hbar \omega_0}{m_e c^2},
\end{equation}
the phase function $\varphi_n(\bm{x})$ can be written as
\begin{eqnarray}
  \varphi_n(\bm{x})&=&(\textbf{k}+k_0\textbf{n})\cdot\textbf{r} - (\omega+\omega_0n_4)t \nonumber\\
   &=&2\pi\left[(\textbf{n}+\textbf{q}/\Omega)\cdot\textbf{r}' - (n_4+q_4/\Omega)X_4\right],\label{phinx}
\end{eqnarray}
where $\textbf{n} = n_1\textbf{e}_1+n_2\textbf{e}_2+n_3\textbf{e}_3 $, $\Omega = \hbar \omega_0/(m_ec^2)$, $\omega_0$ is the frequency of the electromagnetic field, $k_0=\omega_0/c=2\pi/\lambda_0$ is the wave number, $\hbar$ is the Planck constant, $m_e$ is the electron rest mass, $c$ is the speed of light in vacuum.

The evolution operator $E(\bm{x})$ is the major characteristic of the whole family of partial solutions $\Psi(\bm{x})$~(\ref{Exc0}). In particular, it provides a convenient way to calculate mean value $\langle{A}\rangle$ of an operator $A$ with respect to function  $\Psi(\bm{x})$
\begin{equation}\label{meanA}
    \langle{A}\rangle=\frac{a_0^{\dag}A_E a_0}{a_0^{\dag}U_Ea_0},
\end{equation}
where
\begin{equation}\label{AE}
    A_E=\int_0^1dX_1\int_0^1dX_2\int_0^1dX_3\int_0^1dX_4 E^{\dag}(\bm{x})AE(\bm{x}),
\end{equation}
\begin{eqnarray}
    U_E&=&\int_0^1dX_1\int_0^1dX_2\int_0^1dX_3\int_0^1dX_4 E^{\dag}(\bm{x})E(\bm{x})\nonumber\\
    &=&\sum_{n\in Sd}S^{\dag}(n)S(n). \label{UE}
\end{eqnarray}

In the subsequent paper, we will use four finite models of ESTC, designated $p$-models with $p=0,1,2,3$. They differ in level of accuracy, volume of calculations, and field of application. The most simple 0-model with $k_L=\{0\}$ and $\mathcal{L}'=\{n_o\}$ is sufficient to obtain the free space solution from Eq.~(\ref{Exc0}) as the limiting case at vanishing field. In $p$-models with $p>0$, the list $k_L$ begins with zero and contains in order increasing numbers $k$ of all lattices $\mathcal{F}_k$ satisfying the condition $g_{4d}[c_0(8+k)]\leq p$. The set $\mathcal{L}'$ comprises all points $n$ of these lattices, complying with the restriction $n\in \mathcal{L}_{cs}(4,2)$, in particular, 648 points of $\mathcal{F}_0$ (see Sec.~\ref{sec:fractal}). For example, in 1-model we use the list
\begin{equation*}
    k_L=\{0,1,2,3,29,30,31,86,88,331,333,1751,1753\}
\end{equation*}
and the system~(\ref{PkbLC}) containing 998 equations. In $p$-models with $p=2$ and $p=3$, $k_L$ has 69 and 210 members, and the system~(\ref{PkbLC}) consists of 1520 and 2199 equations, respectively.

\section{\label{sec:accur}Evaluating accuracy of solutions}
The distinguishing feature of the presented technique is that each step of the recurrent procedure expands the subsystem of equations for which it provides the exact fundamental solution. One can check the calculation for accuracy by using relations~\cite{ESTCp1}
\begin{equation}\label{rok}
    \rho_k^{\dag}(n)=\rho_k^2(n)=\rho_k(n), \quad tr[\rho_k(n)]=4, \quad n \in \mathcal{L},
\end{equation}
\begin{equation}\label{romn}
    \rho_k(m)\rho_l(n)=0 \text{ if } k\neq l \text{ or (and) } m\neq n.
\end{equation}
In terms of matrices $R_k(m',m,n')$, they can can written as:
\begin{equation}\label{trace}
    \sum_{n\in F_d(k,m)}tr\left[R_k(n,m,n)\right]=4,
\end{equation}
\begin{eqnarray}
 \sum_{p\in F_d(k,m)}&&R_k(m',m,p)R_k(p,m,n') = R_k(m',m,n'),\nonumber\\
                     &&m',n'\in F_d(k,m),  \label{rok2}
\end{eqnarray}
\begin{eqnarray}
  \sum_{p\in F_d(k,m)\cap F_d(l,n)}R_k(m',m,p)R_l(p,n,n') = 0 \label{rokl}\\
  \text{ if } k\neq l \text{ or (and) } m\neq n,\nonumber\\
  \text{ and } m'\in F_d(k,m), n'\in F_d(l,n),\nonumber
\end{eqnarray}
where $F_d(k,m)$  is the subset of $\mathcal{L}$ containing $n'$ with nonzero matrices $\Phi_k(m,n')$ ($F$-domain, see~\cite{ESTCp1}). For the problem under study, the Dirac equation reduces to an infinite system of homogeneous linear equations with matrix coefficients $V(n,s)$~\cite{ESTCp1}. Substitution of Eq.~(\ref{Snc0}) into the left side of these equations reduces it to the form $\mathcal{V}_S(n)a_0$, where
\begin{equation}\label{VSn}
    \mathcal{V}_S(n)=\sum_{s\in S_{13}}V(n,s)S(n+s),
\end{equation}
and $S_{13}$ is set of shifts $s$ with $g_{4d}(s)\leq 1$ [see below the first 13 members of the list $S_{69}$~(\ref{S69})]. At $n\in \mathcal{L}'$, the equation $\mathcal{V}_S(n)a_0=0$ is satisfied at any $a_0$, because in this domain $\mathcal{V}_S(n)\equiv 0$. This provides means for final numerical checking of the fundamental solution $\mathcal{S'}$ of the system (\ref{PkbLC}) and the evolution operator $E(\bm{x})$~(\ref{Ex}) for accuracy.

Let $\mathcal{D}$ be a differential operator in a space $\mathcal{V}_{\Psi}$ of scalar, vector, spinor, or bispinor functions, and $\|\Psi\|$ be the norm of $\Psi$ on $\mathcal{V}_{\Psi}$. The functional
\begin{equation}\label{Rpsi}
    \mathcal{R}: \Psi\mapsto \mathcal{R}[\Psi]=\frac{\|\Psi_D\|}{\|\Psi\|}
\end{equation}
where $\Psi_D=\mathcal{D}\Psi$, evaluates the relative residual at the substitution of $\Psi$ into the differential equation $\mathcal{D}\Psi=0$. It provides a fitness criterion to compare in accuracy various approximate solutions of this equation. For an exact solution $\Psi$, the residual $\Psi_D$ vanishes, i.e., $\mathcal{R}[\Psi]=0$. If $\Psi_D\neq 0$, but $\mathcal{R}[\Psi]\ll 1$, the function $\Psi$ may be treated as a reasonable approximation to the exact solution, and the smaller is $\mathcal{R}[\Psi]$, the more accurate is the approximation. In terms of distances $d=\|\Psi\|$ and $d_D=\|\Psi_D\|$ of $\Psi$ and $\Psi_D$ to the origin of $\mathcal{V}_{\Psi}$ (the zero function), one can graphically describe $\mathcal{R}[\Psi]$  as shrinkage in distance $\mathcal{R}[\Psi]=d_D/d$. The functional $\mathcal{R}$, as applied to a family of functions $\Psi(\bm{x},\lambda)$ with members specified by a parameter $\lambda$, results in function $\mathcal{R}[\Psi(\bm{x},\lambda)]$ of $\lambda$, denoted below $\mathcal{R}(\lambda)$ for short.

To introduce this criterion in the problem under consideration, we first transform the Dirac equation in \cite{ESTCp1} to the equivalent equation $\mathcal{D}\Psi=0$ with the dimensionless operator
\begin{equation}\label{Ddim}
    \mathcal{D}=\sum_{k=1}^3 \alpha_k\left(-\frac{i \hbar}{m_e c}\frac{\partial}{\partial x_k} - A'_k\right) - \frac{i \hbar}{m_e c^2}\frac{\partial}{\partial t} + \alpha_4,
\end{equation}
where $\alpha_j$ are Dirac matrices, and $A'_k$ is defined in \cite{ESTCp1}. From Eqs.~(\ref{Exc0}) and (\ref{Ddim}) follows
\begin{equation}\label{psid}
    \Psi_{D}(\bm{x})=\mathcal{D}\Psi(\bm{x})=D(\bm{x})a_0,
\end{equation}
where
\begin{equation}\label{Dx}
    D(\bm{x})=\mathcal{D}E(\bm{x}) = \sum_{n\in Sd}[D_n - D_A(\bm{x})]e^{i \varphi_n(\bm{x})}S(n)
\end{equation}
is the evolution operator describing the family of remainder functions $\Psi_D$, and
\begin{eqnarray}
  &&D_n=\sum_{k=1}^3 \alpha_k(q_k +n_k\Omega) - U(q_4 +n_4\Omega) + \alpha_4,\\
  &&D_A(\bm{x})=\sum_{k=1}^3 \alpha_k A'_k(\bm{x}).
\end{eqnarray}
The norm of $\Psi_D$ (\ref{psid}) can be written as
\begin{equation}\label{npsid}
    \left\|\Psi_D\right\|=\sqrt{a_0^{\dag} U_Da_0},
\end{equation}
where
\begin{eqnarray}
  U_D=&&\int_0^1dX_1\int_0^1dX_2\int_0^1dX_3\int_0^1dX_4 D^{\dag}(\bm{x})D(\bm{x})\nonumber\\
     =&&\sum_{m,n\in S_d}S^{\dag}(m)\left[D_m D_n\delta(n-m)\right. \label{UD}\\
      &&\left. -\mathcal{A}_1(m,n)D_n - D_m\mathcal{A}_1(m,n) + \mathcal{A}_2(m,n)U \right]S(n),\nonumber
\end{eqnarray}
\begin{eqnarray}
  \mathcal{A}_1(m,n)=\sum_{k=1}^3\alpha_k\sum_{j=1}^6&&\left[A_{jk}\delta(n-m+s_j)\right.\nonumber \\
                     &&\left.+A_{jk}^{\ast}\delta(n-m-s_j)\right],\label{A1mn}
\end{eqnarray}
\begin{eqnarray}
   \mathcal{A}_2(m,n)=\sum_{j,l=1}^6&&\left(\textbf{A}_j\cdot\textbf{A}_l\delta(n-m+s_j+s_l)\right.\nonumber\\
   &&+\textbf{A}_j\cdot\textbf{A}_l^{\ast}\delta(n-m+s_j-s_l)\nonumber\\
   &&+\textbf{A}_j^{\ast}\cdot\textbf{A}_l\delta(n-m-s_j+s_l)\nonumber\\
   &&\left.+\textbf{A}_j^{\ast}\cdot\textbf{A}_l^{\ast}\delta(n-m-s_j-s_l)\right), \label{A2mn}
\end{eqnarray}
\begin{eqnarray*}
  s_1&=&(1,0,0,1),\quad s_2=(0,1,0,1),\\
  s_3&=&(0,0,1,1),\quad s_4=(-1,0,0,1),\\
  s_5&=&(0,-1,0,1),\quad s_6=(0,0,-1,1),
\end{eqnarray*}
vectors $\textbf{A}_j$ and their components $A_{jk}$ are specified in~\cite{ESTCp1}. Thus, for the function $\Psi$~(\ref{Exc0}), from the definition~(\ref{Rpsi}) follows
\begin{equation}\label{Rc0}
   \mathcal{R}=\sqrt{\frac{a_0^{\dag} U_Da_0}{a_0^{\dag} U_Ea_0}},
\end{equation}
where $U_E$ and $U_D$ given by Eqs.~(\ref{UE}) and (\ref{UD}), respectively.

\section{Conclusion}
The projection operator $\mathcal{S'}$~(\ref{SkbL}) defines the exact fundamental solution of the finite subsystem~(\ref{PkbLC}) which expands with each new step of the recurrent process. The relations presented above form the complete set which is sufficient for the fractal expansion of this subsystem to a finite model of ESTC of any desired size. A criterion for evaluating accuracy of the approximate solutions, obtained by the use of such model, is suggested. It plays a leading role in search for best approximate solutions in the framework of the selected model. The corresponding examples will be presented in the subsequent paper.

\appendix*
\section{}
In this series of papers, we intensively use indexing of various mathematical objects by points $n=(n_1,n_2,n_3,n_4)$ of the integer lattice $\mathcal{L}$ with even values of the sum $n_1+n_2+n_3+n_4$. The introduced below sequential numbering of these points drastically simplifies both numerical implementation of the presented fractal technique and analysis of solutions, because it takes into account the specific Fourier spectra of the electromagnetic field of ESTC and the wave function, as well the structure of the finite models of ESTCs described above. It is of particular assistance in the analysis of partial solutions with the localized amplitude $C_0$~(\ref{C0n0}).

Let us define functions $g_{3d}(n)$ and $g_{4d}(n)$ of $n=(n_1,n_2,n_3,n_4)\in \mathcal{L}$ as follows
\begin{equation}\label{gen3d}
   g_{3d}(n)\equiv g_{3d}(n_1,n_2,n_3,n_4)=|n_1|+|n_2|+|n_3|,
\end{equation}
\begin{eqnarray}\label{gen}
    g_{4d}(n)&\equiv &g_{4d}(n_1,n_2,n_3,n_4)\nonumber\\
    &=&\max\{|n_1|+|n_2|+|n_3|,|n_4|\}.
\end{eqnarray}
For any $n\in \mathcal{L}$, integers $g_{3d}(n)$ and $g_{4d}(n)$ have the same parity. First we split $\mathcal{L}$ into the infinite sequence of finite subsets $\mathcal{G}^{\{p\}}$ ($p$--generations) composed of all $n\in \mathcal{L}$ with $g_{4d}(n)=p, p=0,1,2,\dots$. Next we split $\mathcal{G}^{\{p\}}$ into subsets $\mathcal{G}^{\{p,r\}}$ composed of members $n$ with $g_{3d}(n)=r\leq p$. Then we split $\mathcal{G}^{\{p,r\}}$ into subsets $\mathcal{G}^{\{p,r,n_4\}}$ of members $n=(n_1,n_2,n_3,n_4)$ with equal values of $n_4$. Finally, we split $\mathcal{G}^{\{p,r,n_4\}}$ into subsets $\mathcal{G}^{\{p,r,n_4,n_3\}}$ of members with equal values of $n_3$. These inclusion relations can be written as
\begin{equation*}
    n\in \mathcal{G}^{\{p,r,n_4,n_3\}}\subset\mathcal{G}^{\{p,r,n_4\}}\subset\mathcal{G}^{\{p,r\}}\subset\mathcal{G}^{\{p\}}\subset\mathcal{L}.
\end{equation*}

\begin{figure}
\includegraphics{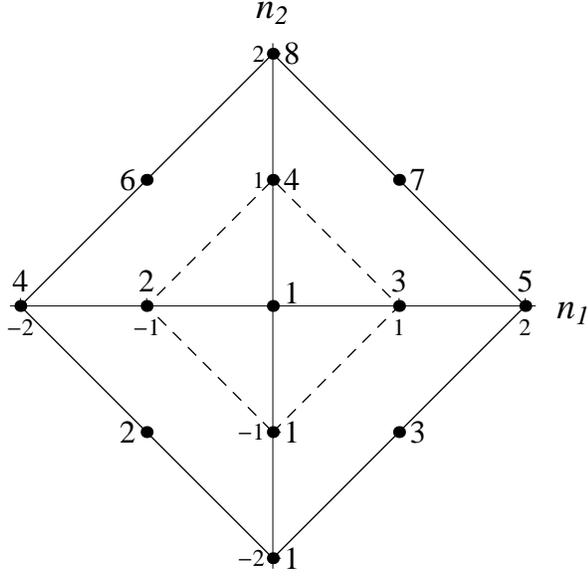}%
\caption{\label{p2fig1}Numbers $i_4$(\ref{i4cases}) at $R=0$ (the central point), $R=1$ (4 points connected by the dash lines) and $R=2$ (8 points connected by the solid lines.}
\end{figure}

To introduce a sequential numbering $i=0,1,2,\dots$ of points $n\in \mathcal{L}$ in the direction of increasing $p=g_{4d}(n)$, we first assign the global number $i=0$ to the single member $n_o=(0,0,0,0)$ of $\mathcal{G}^{\{0\}}\equiv \{n_o\}$ and local numbers $i_4=1,\dots$ to members of $\mathcal{G}^{\{p,r,n_4,n_3\}}$, which differ from one another only by values of $n_1$ and $n_2$, as follows
\begin{eqnarray}
    i_4&=&1 \text{ for } n_1=0 \text{ and } n_2\leq 0,\nonumber\\
       &=&2(R+n_2) \text{ for } n_1<0,\nonumber\\
       &=&2(R+n_2)+1 \text{ for } n_1>0,\nonumber\\
       &=&4n_2 \text{ for } n_1=0 \text{ and } n_2>0,\label{i4cases}
\end{eqnarray}
where $R=|n_1|+|n_2|=r-|n_3|$, see also Fig.~\ref{p2fig1}. The total number of $\mathcal{G}^{\{p,r,n_4,n_3\}}$ members depend on $R$ as
\begin{eqnarray}
    N_4(R)&=&1 \text{ for } R=0,\nonumber\\
          &=&4R \text{ for } R>0,\label{N4R}
\end{eqnarray}
and $i_4=1,\dots,N_4(R)$.

Next we introduce local numbers $i_3$ of $\mathcal{G}^{\{p,r,n_4\}}$ members as
\begin{equation}\label{i3}
    i_3=M_4(r,n_3-1)+i_4,
\end{equation}
where
\begin{eqnarray}
  M_4&&(r,n_3)=\sum_{n'_3=-r}^{n_3}N_4(r -|n'_3|) \label{M4cases}\\
            &&=0 \text{ for } n_3<-r, \nonumber\\
            &&=1+2(n_3+r)(n_3+r+1) \text{ for } -r\leq n_3\leq 0,\nonumber\\
            &&=1+2r(r+1)-2n_3(n_3+1-2r) \text{ for } 0<n_3<r,\nonumber\\
            &&=2+4r^2 \text{ for } n_3=r>0,\nonumber
\end{eqnarray}
and $i_4$ is defined by Eq.~(\ref{i4cases}). The total number of $\mathcal{G}^{\{p,r,n_4\}}$ members is
\begin{eqnarray}
  N_3(r)=M_4(r,r)&=&1 \text{ for } r=0,\nonumber\\
                   &=&2+4r^2 \text{ for } r>0,\label{N3r}
\end{eqnarray}
and $i_3=1,\dots,N_3(r)$. One can visualize $\mathcal{G}^{\{p,r,n_4\}}$ as a set of elements $n=(n_1,n_2,n_3,n_4)$ with projections $(n_1,n_2,n_3)$ onto the three--dimensional space, lying in the eight faces of the regular octahedron with the six corner points $(\pm r,0,0), (0,\pm r,0), (0,0,\pm r)$.

To enumerate elements of $\mathcal{G}^{\{p,r\}}$, we specify the numeration order of its subsets $\mathcal{G}^{\{p,r,n_4\}}$ by $j_4=1,\dots,j_{4max}$,where
\begin{eqnarray}
  j_4&=&1 \text{ for } r<p=|n_4| \text{ and } n_4<0,\nonumber\\
     &=&2 \text{ for } r<p=|n_4| \text{ and } n_4>0,\nonumber\\
     &=&-n_4 \text{ for } r=p \text{ and } n_4<0,\nonumber\\
     &=&1+n_4 \text{ for } r=p \text{ and } n_4\geq 0, \label{j4}
\end{eqnarray}
\begin{eqnarray}
\label{j4max}
    j_{4max}&=&2 \text{ for } r<p\nonumber\\
            &=&1+p \text{ for } r=p.
\end{eqnarray}
All these subsets have the same total number of members $N_3(r)$, so that
\begin{eqnarray}
  N_2(p,r)&=&2 N_3(r) \text{ for } r<p,\nonumber\\
          &=&(1+p)N_3(p) \text{ for } r=p \label{N2}
\end{eqnarray}
is the total number of $\mathcal{G}^{\{p,r\}}$ members enumerated by
\begin{equation}\label{i2}
    i_2=(j_4-1)N_3(r)+i_3,
\end{equation}
where $i_3$ is given by Eq.~(\ref{i3}).

At any given $p$, $p$--generation $\mathcal{G}^{\{p\}}$  consists of subsets $\mathcal{G}^{\{p,r\}}$, where $r$ has the same parity as $p$ and takes
\begin{equation}\label{kmax}
    k_{max}=\frac{p}{2}+\frac{3+(-1)^p}{4}
\end{equation}
different values from $r_{min}=[1-(-1)^p]/2$ to $r_{max}=p$. The total number of elements of all subsets $\mathcal{G}^{\{p,r'\}}\subset\mathcal{G}^{\{p\}}$ with $r'\leq r\leq p$ is given by
\begin{eqnarray}
  M_2(p,r)&=&\sum_{j=1}^k N_2[p,p-2(k_{max}-j)]\label{M2cases}\\
          &=&0 \text{ for } r<r_{min},\nonumber\\
          &=&\frac23 (r+1)(2r^2+4r+3) \text{ for } r<p,\nonumber\\
          &=&1 \text{ for } r=p=0,\nonumber\\
          &=&\frac43 p(4p^2+5) \text{ for } r=p>0,\nonumber
\end{eqnarray}
where $k=r/2+[3+(-1)^p]/4$. The total number of $p$-generation members is $N_1(p)\equiv M_2(p,p)$, and these members are enumerated by
\begin{equation}\label{eqi1}
    i_1=M_2(p,r-2)+i_2,
\end{equation}
where $i_2$ is given by Eq.~(\ref{i2}).

Finally, we introduce the global numbering [$n=(n_1,n_2,n_3,n_4)\in \mathcal{L}\mapsto i$] of the lattice $\mathcal{L}$ points as
\begin{eqnarray}
   i=&&M_1(p-1)+i_1=M_1(p-1)+M_2(p,r-2) \nonumber\\
     &&+(j_4-1)N_3(r)+M_4(r,n_3-1)+i_4,
\end{eqnarray}
where
\begin{equation}\label{M1}
    M_1(p)=\sum_{k=1}^p N_1(p)=\frac23 p(p+1)(2p^2+2p+5)
\end{equation}
is the global number of the last element of $p$-generation ordered as described above.

With this numeration, $\mathcal{L}$ becomes the ordered infite set and the inverse mapping $s: i\mapsto s_h(i)=(n_1,n_2,n_3,n_4)$ is defined as follows. The number $i=0$ defines $n_o=(0,0,0,0)$. At $i>0$, we first find the generation number $p$ from the condition
\begin{equation}\label{pbyi}
    M_1(p-1)<i\leq M_1(p)
\end{equation}
and calculate the local number
\begin{equation}\label{i1byi}
    i_1=i-M_1(p-1).
\end{equation}
Next we determine $r$ and $i_2$ from the relations
\begin{eqnarray}
  M_2(p,r-2)<i_1\leq M_2(p,r) \\
  i_2=i_1-M_2(p,r-2).
\end{eqnarray}
The relations (\ref{j4}), (\ref{j4max}) and (\ref{i2}) make it possible first to find $j_4$ from the condition
\begin{equation}\label{j4by}
    j_4-1<\frac{i_2}{N_3(r)}\leq j_4
\end{equation}
and then
\begin{eqnarray}
  n_4&=&(-1)^{j_4}p \text{ for } r<p,\label{n4by}\\
     &=&(-1)^{p+j_4+1}j_4 + \left[(-1)^{p+j_4}-1\right]/2 \text{ for } r=p,\nonumber
\end{eqnarray}
\begin{equation}\label{i3by}
    i_3=i_2-(j_4 - 1)N_3(r).
\end{equation}
Thereafter we find $n_3$ and $i_4$:
\begin{eqnarray}
  M_4(r,n_3-1)<i_3\leq M_3(r,n_3),\\
  i_4=i_3 - M_4(r,n_3-1).
\end{eqnarray}
Finally, we obtain the last two components
\begin{eqnarray}
  n_2 &=&  -R+i_4/2+ \left[(-1)^{i_4}-1\right]/4,\\
  n_1 &=& (-1)^{i_4}(|n_2|-R),
\end{eqnarray}
where $R=r-|n_3|$.

As an illustration let us consider the values of the function $s_h(i)$ at $i=0,1\dots,68$, which define all members of $p$-generations with $p=0,1,2$. The total number $N_3(r)$~(\ref{N3r}) of the members of the set $\mathcal{G}^{\{p,r,n_4\}}$ depends only on $r$, in particular, $N_3(0)=1, N_3(1)=6$, and $N_3(2)=18$. The first value $s_h(0)=n_o=(0,0,0,0)$ is the single member of the set $\mathcal{G}^{\{0,0,0\}}\equiv \mathcal{G}^{\{0\}}\equiv \{n_o\}$. The following six values [${s_h(i),i=1,\dots,6}$] are the members of the set $\mathcal{G}^{\{1,1,-1\}}$, whereas the next six members [${s_h(i),i=7,\dots,12}$] are members of $\mathcal{G}^{\{1,1,1\}}$. Each of the sets $\mathcal{G}^{\{2,0,-2\}}$ and $\mathcal{G}^{\{2,0,2\}}$ has only one member: $s(13)=(0,0,0,-2)$, and $s(14)=(0,0,0,2)$, respectively. At $i=15,\dots,32; i=33,\dots,50$; and $i=51,\dots,68$, the function $s_h(i)$ gives the members of $\mathcal{G}^{\{2,2,0\}}$, $\mathcal{G}^{\{2,2,-2\}}$, and $\mathcal{G}^{\{2,2,2\}}$, respectively. Consequently, the list $S_{69}$ of the first 69 values of the function $s_h(i)$, which contains members of $p$-generations with $p=0,1,2$ has the following form:
\begin{eqnarray}
 S_{69}=&&\left\{(0, 0, 0, 0), \right. \label{S69}\\
 &&(0, 0, -1, -1),(0, -1, 0, -1),(-1, 0, 0, -1),\nonumber\\
 &&(1, 0, 0, -1),(0, 1, 0, -1),(0, 0, 1, -1),\nonumber\\
 &&(0, 0, -1, 1),(0, -1, 0, 1),(-1, 0, 0, 1),\nonumber\\
 &&(1, 0, 0, 1),(0, 1, 0, 1),(0, 0, 1, 1),\nonumber\\
 &&(0, 0, 0, -2),(0, 0, 0, 2),\nonumber\\
 &&(0, 0, -2, 0),(0, -1, -1, 0),(-1, 0, -1, 0),\nonumber\\
 &&(1, 0, -1, 0),(0, 1, -1, 0),(0, -2, 0, 0),\nonumber\\
 &&(-1, -1, 0, 0),(1, -1, 0, 0),(-2, 0, 0, 0),\nonumber\\
 &&(2, 0, 0, 0),(-1, 1, 0, 0),(1, 1, 0, 0),\nonumber\\
 &&(0, 2, 0, 0),(0, -1, 1, 0),(-1, 0, 1, 0),\nonumber\\
 &&(1, 0, 1, 0),(0, 1, 1, 0),(0, 0, 2, 0),\nonumber\\
 &&(0, 0, -2, -2),(0, -1, -1, -2),(-1, 0, -1, -2),\nonumber\\
 &&(1, 0, -1, -2),(0, 1, -1, -2),(0, -2, 0, -2),\nonumber\\
 &&(-1, -1, 0, -2),(1, -1, 0, -2),(-2, 0, 0, -2),\nonumber\\
 &&(2, 0, 0, -2),(-1, 1, 0, -2),(1, 1, 0, -2),\nonumber\\
 &&(0, 2, 0, -2),(0, -1, 1, -2),(-1, 0, 1, -2),\nonumber\\
 &&(1, 0, 1, -2),(0, 1, 1, -2),(0, 0, 2, -2),\nonumber\\
 &&(0, 0, -2, 2),(0, -1, -1, 2),(-1, 0, -1, 2),\nonumber\\
 &&(1, 0, -1, 2),(0, 1, -1, 2),(0, -2, 0, 2),\nonumber\\
 &&(-1, -1, 0, 2),(1, -1, 0, 2),(-2, 0, 0, 2),\nonumber\\
 &&(2, 0, 0, 2),(-1, 1, 0, 2),(1, 1, 0, 2),\nonumber\\
 &&(0, 2, 0, 2),(0, -1, 1, 2), (-1, 0, 1, 2),\nonumber\\
 &&\left. (1, 0, 1, 2),(0, 1, 1, 2), (0, 0, 2, 2)\right\}. \nonumber
\end{eqnarray}

\bibliography{Borzdov1}
\end{document}